\documentclass{iaus}
\usepackage{graphicx}
\usepackage{subfigure}
\usepackage{amsmath}
\usepackage{natbib}
\usepackage{bm}

\title[Turbulence and magnetic spots at the surface of hot massive stars]  
{Turbulence and magnetic spots at the surface of hot massive stars}

\author[Matteo  Cantiello et al.]   
{Matteo  Cantiello$^1$, Jonathan Braithwaite$^1$, Axel Brandenburg$^{2,3}$, Fabio Del Sordo$^{2,3}$, Petri K\"apyl\"a$^{2,4}$, Norbert Langer$^1$}
 
\affiliation{$^1$Argelander-Institut f\"ur Astronomie der Universit\"at Bonn, Auf dem H\"ugel 71, D--53121 Bonn, Germany
 email: {\tt cantiello@astro.uni-bonn.de};
$^2$NORDITA, AlbaNova University Center, Roslagstullsbacken 23, SE-10691 Stockholm, Sweden;
$^3$Department of Astronomy, AlbaNova University Center,
Stockholm University, SE--10691 Stockholm, Sweden;
$^4$Department of Physics, Gustaf H\"allstr\"omin katu 2a (PO Box 64), FI-00014, University of Helsinki, Finland}
 
\pubyear{2010}
\volume{273}   
\pagerange{}
\jname{Physics of Sun and Star Spots}
\editors{}

 \def\simle{\mathrel{\hbox{\rlap{\hbox{\lower4pt\hbox{$\sim$}}}\hbox{$<$}}}}
 \def\simgr{\mathrel{\hbox{\rlap{\hbox{\lower4pt\hbox{$\sim$}}}\hbox{$>$}}}}

 \def\c2{^{12}{\rm C}}
 \def\c3{^{13}{\rm C}}
 \def\n14{^{14}{\rm N}}
 \def\c1213{^{12}{\rm C}/^{13}{\rm C}}
 \def\he3he4{^3{\rm He}/^4{\rm He}}

\newcommand{\EQ}{\begin{equation}}
\newcommand{\EE}{\end{equation}}
\newcommand{\EQA}{\begin{eqnarray}}
\newcommand{\EEA}{\end{eqnarray}}

\begin{document}
\maketitle

\begin{abstract}
Hot luminous stars show a variety of phenomena in their photospheres and in their winds
%MC: interpretations -> explanations
which still lack clear physical explanations at this time. Among these phenomena are
non-thermal line broadening, line profile variability (LPVs), discrete absorption components (DACs), wind
clumping and stochastically excited pulsations. \citet{2009A&A...499..279C} argued that a convection zone close to the surface of hot, massive stars, could be responsible for some of these phenomena.  This convective zone is caused by a peak in the opacity due to iron recombination and for this reason is referred to as the ``iron convection zone'' (FeCZ).
3D MHD simulations are used to explore the possible effects of such subsurface convection  on the surface properties of hot, massive stars.
We argue that turbulence and localized magnetic spots at the surface are the likely consequence of subsurface convection in early type stars. 

\end{abstract}
\section{Introduction}
During their main sequence evolution, massive stars can develop convective regions very close to  their surface.
These regions are caused by an opacity peak associated with iron ionization. 
\citet{2009A&A...499..279C} found a correlation between the occurrence and properties of  subsurface convection, and microturbulence at the surface of hot massive stars.
This correlation has been recently corroborated by new observations of microturbulence in massive stars \citep{2010MNRAS.404.1306F}. Moreover there is growing evidence that the FeCZ is responsible for the observed solar-like oscillations  at the surface of OB stars \citep{2009Sci...324.1540B,2010A&A...519A..38D}. These observations seem to confirm the occurrence of such a convective region  and its importance for the surface properties of early type stars.
\section{3D MHD Simulations of subsurface convection}\label{mhd}
The transport of energy by convection in the FeCZ is relatively inefficient. Radiation dominates and transports more than 95\% of the total flux.
The convective layer is very close to the photosphere, above which strong winds are accelerated. 
In rotating stars, the associated angular momentum loss might also drive strong differential rotation in these layers.

We perform 3D MHD simulations of the FeCZ.
We use a setup similar to the one of \citet{2008A&A...491..353K}. This is described in more detail in  \citet{2010arXiv1009.4462C}.  
As a preliminary study we perform simulations with modest resolution, where the density contrast between the bottom of the convective layer and the top of the domain is only $\sim$20.  This is about ten times smaller than in the case of the FeCZ. Moreover, the ratio of the convective to radiative flux is about 0.3,  higher than in the FeCZ case. 
Therefore, at this stage, the velocities of convective motions cannot 
be directly compared with those of more realistic models, even though they
still use mixing length theory.
However, already in these preliminary runs we could follow the excitation and propagation of gravity waves above the convective region. 
Energy is transported up to the top layer by gravity waves, where the maximum of the energy is deposited at those wavelengths that are
resonant with the scale of convective motions, as predicted, for example, by \citet{gk90}.
\citet{2008A&A...491..353K}  found excitation of a large scale dynamo in simulations of turbulent convection including rotation and shear.
Our computational setup is very similar, so it is not surprising that we confirm this result.
Dynamo action reaching equipartition is found in our simulations that include shear and rotation, with magnetic fields on scales  larger than the scale of convection.
\begin{figure}[b]
\begin{center}
 \includegraphics[width=6.1in]{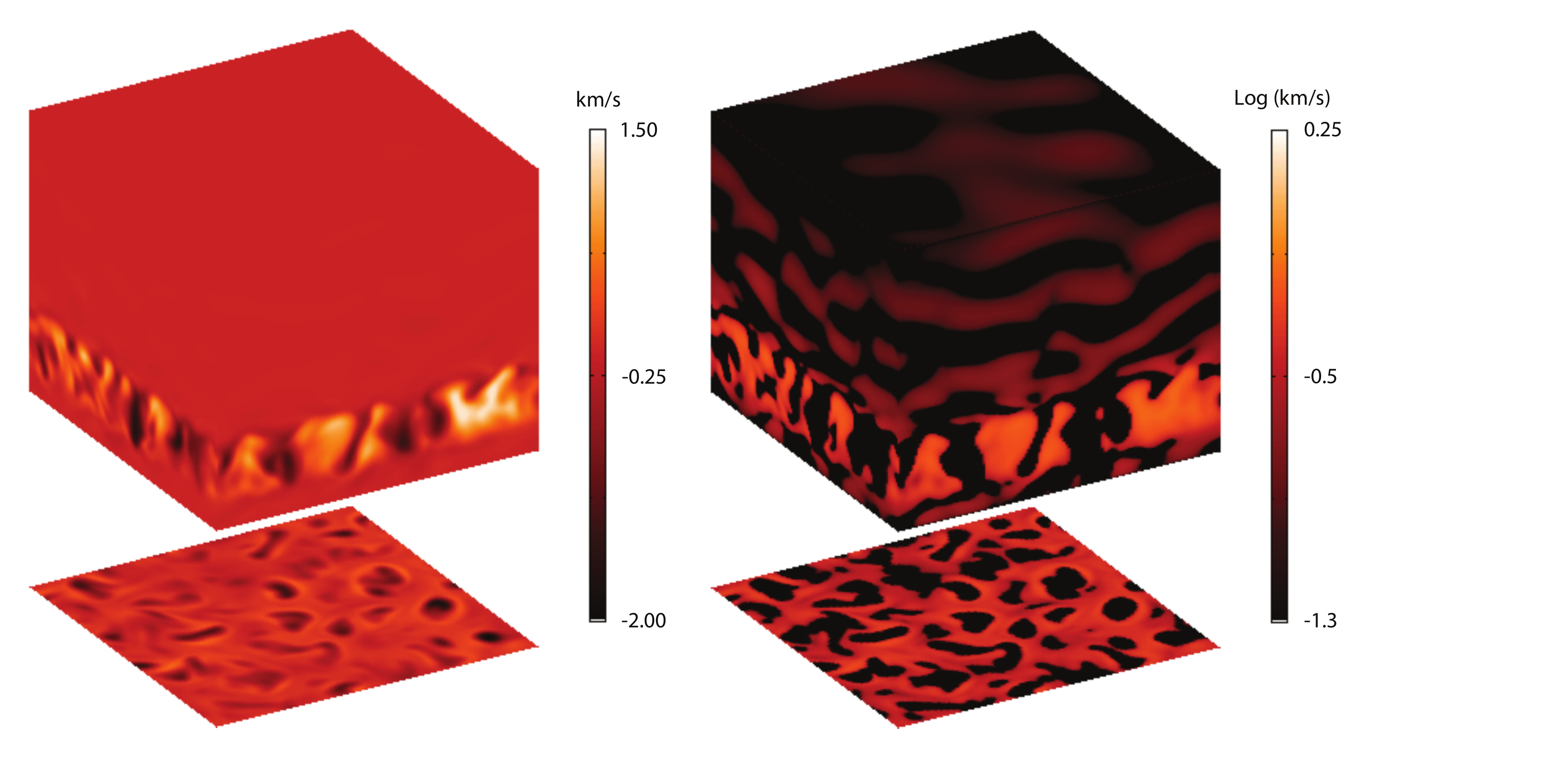} 
 %MC: slightly modified the caption to explain a little better the right plot.
 \caption{Simulation of subsurface convection. Starting from the top, the computational box is divided into
three layers: a radiative layer with an upper cooling boundary, a convectively unstable layer
and another stable layer at the bottom.  The left panel shows a snapshot of the vertical velocity field in the simulation.  In the right panel we show, for the same calculation, the logarithm of the vertical velocity field with $v_z > 0$. The plane below each box show
the vertical velocity field at the lower boundary of the convective layer. The right box shows the gravity waves propagating in the radiative layer. 
Due to the low resolution of these preliminary runs, the amplitude of the convective velocities and of gravity waves can not be directly compared to the stellar case.}
   \label{fig1}
\end{center}
\end{figure}
\section{Surface turbulence}
Microturbulence measures the amplitude of plasma motions that are of non-thermal origin and have a correlation length smaller than the region of line formation. In spectroscopy the microturbulence parameter needs to be estimated  to derive consistent surface abundances for one element from different photospheric absorption lines through stellar model atmospheres
\citep[among many others][]{1996A&A...315...95R,1998A&A...332..681H}. It is unfortunate that so far microturbulence has always been used as a fudge-factor, as its physical origin is not understood.  However \citet{2009A&A...499..279C} suggested that  the presence of convection below the surface of hot massive stars could explain turbulence in their photospheres, as measured by the microturbulence parameter. In fact  a clear correlation has been found between the presence and properties of subsurface convection and the amplitude of microturbulent velocities in the photospheres of early type stars.  Using our 3D MHD simulations we can study the excitation of gravity waves and their propagation. Such waves are excited in the convection zone and, once they reach the surface, they could produce the observed microturbulent velocity field. 
Further work still needs to be done, 
but ideally our simulations might lead to a situation in which the microturbulence is no longer a fudge-parameter, but a function of the stellar parameters. 
\section{Magnetic spots}
The occurrence of convection zones close to the surface of hot massive stars opens
a new scenario: If dynamo action is excited in the FeCZ, magnetic fields can be readily produced in the envelopes of OB stars. 
Such magnetic fields could reach the surface due to magnetic buoyancy.  
Following \citet{2009A&A...499..279C} and supported by our 3D MHD simulations, we can assume that magnetic fields at equipartition level are generated in the FeCZ.
This means that the magnetic energy density is equal to the  kinetic energy density, giving amplitudes up to 3kG. 
Such fields may reach the stellar surface and result in localized magnetic spots. Details of how the magnetic fields produced in the subsurface convection may reach the stellar surface and with which amplitude, will be discussed in a forthcoming paper (Cantiello \& Braithwaite, in prep.).
Such magnetic spots, if they exist, could have remarkable effects on observable properties of early type stars. 
Surface magnetic fields have been linked to several observed phenomena 
in OB stars, e.g. discrete absorption components 
(DACs) in UV resonance lines
\citep[e.g.,][]{1988MNRAS.233..123P,1995ApJ...452L..53M,1997A&A...327..281K,2002A&A...388..587P}, 
which are thought to diagnose large scale coherent wind
anisotropies \citep{1996ApJ...462..469C,2008ApJ...678..408L},
or the less coherent line profile variability  \citep{1996ApJS..103..475F,1997A&A...327..699F}.
Also non-thermal X-ray emission of OB main sequence stars has been proposed to relate
to surface magnetic fields \citep[e.g.,][]{1997ApJ...485L..29B,2002ApJ...576..413U}.
Magnetic fields generated in subsurface convective zones  could affect not only the stellar wind mass loss, but also the associated angular momentum loss from the star.
This could have important consequences for the evolution of massive stars.
\section{Discussion}
An intriguing connection between  the presence of sub-photospheric convective motions and microturbulence in early-type stars has been found by \citet{2009A&A...499..279C}. 
A picture in which the FeCZ influences surface properties of OB stars is supported also by the recent discovery of solar-like oscillations in early type stars \citep{2009Sci...324.1540B,2010A&A...519A..38D} and new measurements of microturbulence \citep{2010MNRAS.404.1306F}. 

We perform 3D MHD simulations of convection to investigate the
excitation and propagation of gravity waves above a subsurface
convection zone. Analytical predictions of \citet{gk90} on the spatial
scale at which the maximum of energy is injected in gravity waves seem
to be confirmed by our preliminary calculations. Further investigation
is required in order to understand if the subsurface convection
expected in OB stars excites gravity waves of the required amplitude
to explain the observed microturbulence in massive stars. In
particular we need higher resolution to increase the Reynolds number
of our simulations and to be able to decrease the ratio of convective to
radiative flux, which is an important parameter in determining the
convective velocities \citep{2005AN....326..681B}.

Simulations of turbulent  convection in the presence of rotation and shear, show dynamo action with magnetic fields reaching equipartition  (K\"apyl\"a et al. 2008). Since massive stars are usually fast rotators,  perhaps the interplay between convection, rotation and shear is able to drive a dynamo in OB stars.  Indeed our simulations of subsurface convection including rotation and shear show  dynamo-generated magnetic fields with equipartition values. This means that fields of $\sim$kG could be present in the FeCZ.  
These magnetic fields might experience buoyant rise and reach the surface of OB stars, where they could have important observational consequences.  In particular it has already been suggested that the discrete absorption components observed in UV lines of massive stars could be produced by low amplitude, small scale magnetic fields at the stellar surface \citep{1994Ap&SS.221..115K}. We will discuss the emergence and appearance of localized magnetic spots at the surface of hot massive stars in a forthcoming paper (Cantiello \& Braithwaite, in prep.).

\bibliographystyle{aa} 
\bibliography{ref}
\end{document}